\documentclass[conference]{IEEEtran}
\IEEEoverridecommandlockouts

\usepackage{cite}
\usepackage{amsmath,amssymb,amsfonts}
\usepackage{algorithmic}
\usepackage{graphicx}
\usepackage{textcomp}
\usepackage{xcolor}

\def\BibTeX{{\rm B\kern-.05em{\sc i\kern-.025em b}\kern-.08em
    T\kern-.1667em\lower.7ex\hbox{E}\kern-.125emX}}

\usepackage{makecell}
\usepackage{multirow}

\usepackage{hyperref} 

\newcolumntype{F}[1]{%
    >{\raggedright\arraybackslash\hspace{0pt}}p{#1}}%
\newcolumntype{T}[1]{%
    >{\centering\arraybackslash\hspace{0pt}}p{#1}}%

\usepackage{caption}
\usepackage{array}

\newif\ifcameraready
\camerareadytrue
    
\begin{document}

\title{Handling Numeric Expressions\\in Automatic Speech Recognition}

\ifcameraready
\author{\IEEEauthorblockN{Christian Huber}
\IEEEauthorblockA{\textit{Interactive Systems Lab}\\
\textit{Karlsruhe Institute of Technology}\\
Karlsruhe, Germany\\
christian.huber@kit.edu}
\and
\IEEEauthorblockN{Alexander Waibel}
\IEEEauthorblockA{\textit{Interactive Systems Lab}\\
\textit{Carnegie Mellon University}\\
Pittsburgh PA, USA\\
waibel@cs.cmu.edu}
}
\else
\author{\IEEEauthorblockN{BLIND}
}
\fi

\maketitle

\begin{abstract}
This paper addresses the problem of correctly formatting numeric expressions in automatic speech recognition (ASR) transcripts. This is challenging since the expected transcript format depends on the context, e.g., 1945 (year) vs. 19:45 (timestamp). 
We compare cascaded and end-to-end approaches to recognize and format numeric expressions such as years, timestamps, currency amounts, and quantities.
For the end-to-end approach, we employed a data generation strategy using a large language model (LLM) together with a text to speech (TTS) model to generate adaptation data.
The results on our test data set show that while approaches based on LLMs perform well in recognizing formatted numeric expressions, adapted end-to-end models offer competitive performance with the advantage of lower latency and inference cost.
\end{abstract}


\section{Introduction}

In the last decade, ASR systems improved tremendously in terms of word error rate (WER) due to more data, more computing power, and better architectures \cite{vaswani2017attention,pham2019very,radford2023robust}.
These systems are normally trained with labeled ASR data, that is, human-transcribed speech or human-correct automatically transcribed speech.

The way in which numeric expressions are transcribed - using numeric literals, e.g., 1945, or number words, e.g., nineteen forty-five - can vary between different datasets or sometimes even within a dataset.
Depending on the usage of the ASR system, different transcript formats might be preferred. For example, when a video conference call is automatically subtitled using an ASR system, the readers might prefer numeric literals, since they are shorter and easier to read. Furthermore, proper interpretation of a numeric expression is critical for downstream processing such as in speech translation \cite{waibel2012simultaneous, fugen2007simultaneous} and summarization \cite{retkowski2023current}. For example, the time of day 18:45 (German) might have to be translated into "sept heures moins quart" in French (literally: "seven hours minus a quarter").

On the other hand, transcripts of an ASR system containing numeric expressions 
should be formatted according to the context in which the numeric expressions occur.
For example, the number word nineteen forty-five should be formatted as 1945, 19:45, \$19.45 or 1,945 if it represents a year, timestamp, currency amount or quantity, respectively.

Often numeric expression formatting is not reflected in the WER because numeric expression formats are normalized before calculating the WER. However, proper formatting of numeric expressions is important because it heavily improves 
readability of the transcript.

Therefore, in this work, we tackle the problem of properly formatting numeric expressions in ASR transcripts. For this, we 1) created a test set containing the numeric expression types year, timestamp, currency amount, and quantity, 2) propose a strategy using an LLM together with a TTS model to obtain synthetic data with which an end-to-end ASR system can be adapted (see Figure \ref{fig:augmentation} and Section \ref{sec:data}), and 3) compared cascaded and end-to-end approaches to recognize the numeric expression types (see Sections \ref{sec:approach} and \ref{seq:res}). 
We show that while approaches based on LLMs perform well in recognizing formatted numeric expressions, adapted end-to-end models offer competitive performance with the advantage of lower latency and inference cost.

\section{Related Work}

\begin{table}[t]
\centering
\begin{tabular}{|T{0.13\textwidth}|T{0.155\textwidth}|T{0.095\textwidth}|}
\hline
Numeric expression type & Number words & Wanted formatting\\
\hline
\hline
Year & \makecell{in nineteen\\ forty-five} & in 1945\\
\hline
Timestamp & \makecell{at quarter to eight\\ (in the evening)} & at 19:45\\
\hline
\makecell{Currency amount\\ (English)} & \makecell{one thousand dollars\\ and fifty cents} & \$1,000.50\\
\hline
\makecell{Currency amount\\ (German)} & \makecell{eintausend Euro\\ und fünfzig Cent} & 1.000,50€\\
\hline
\makecell{Quantity\\ (English)} & \makecell{two thousand pieces} & \makecell{2,000 pieces}\\
\hline
\makecell{Quantity\\ (German)} & \makecell{zweitausend Teile} & \makecell{2.000 Teile}\\
\hline
\end{tabular}
\vskip 5pt
\captionsetup{width=0.86\linewidth}
\caption{Examples of wanted formatting of numeric expressions for different numeric expression types.}
\vskip -5pt
\label{tab:formatting}
\end{table}

Until a few years ago, most ASR systems were trained to output lowercase transcripts without punctuation \cite{nguyen2020improving}.
For this, the transcripts of the training data were normalized.
To obtain a transcript that contains casing and punctuation, inverse text normalization \cite{shugrina2010formatting} was applied. This was done by applying a text segmentation model \cite{cho2012segmentation,cho2017nmt} after the transcription.
For the text segmentation step auto-regressive models similar to models used in machine translation can be used.
To minimize the training-test mismatch in the input distribution, such a text segmentation model should be trained on
hypotheses specific to some ASR model.
Therefore, when the ASR model is changed, the text segmentation model should also be changed.

The recently introduced LLMs \cite{brown2020language,touvron2023llama,touvron2023llama2} can also be used as a text segmentation model, i.e., to reformat the ASR hypotheses.
LLMs are pre-trained on a lot of text to predict the next token and then adapted to follow prompt instructions. In-context learning \cite{dong2022survey} can be used to increase performance without changing the weights of the LLM by providing examples on how prompt instructions should be answered.

On the other hand, ASR systems have recently moved more and more toward end-to-end approaches where the transcript already contains casing and punctuation \cite{radford2023robust}. This has the advantage that only one model has to be executed, decreasing latency and reducing maintenance effort.
Furthermore, end-to-end approaches search for a global optimum and with enough training data, this works well \cite{radford2023robust}.
The drawback is that the formatting of numeric expressions in the transcript can not be easily changed with text-only data, and the question is how to get ASR data with suitable numeric expression formatting. 

We use a TTS model for synthetic data generation (see next section). Other works \cite{rossenbach2020generating,fazel2021synthasr} have shown that it is possible to use synthetic TTS data to improve ASR performance. 
Furthermore, using synthetic TTS data can improve robustness by providing a controlled and unbiased data set. Unlike real-world speech, which may contain an imbalanced distribution of certain numeric expressions (e.g., 90\% appearing far more frequently than 47\% \cite{ardila2019common}), a TTS model can generate balanced examples. This can help to recognize less common words or patterns, such as 47\%, with greater accuracy, thus reducing recognition bias.

\section{Experiments}
\label{sec:experiments}

\subsection{Data}
\label{sec:data}

\begin{figure}[t]
  \centering
  \boxed{
  \includegraphics[trim=7.3cm 6.8cm 8.2cm 6.0cm,clip,width=0.9\columnwidth]{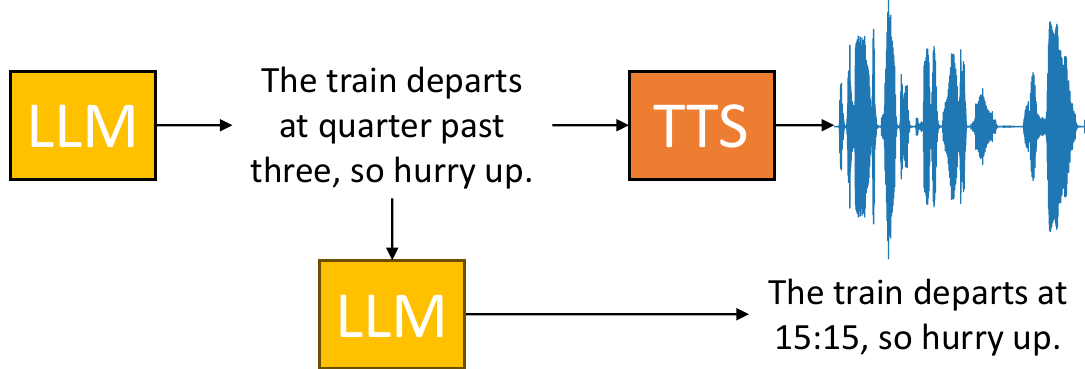}
  }
  \vskip 5pt
  \captionsetup{width=0.93\linewidth}
  \caption{Data generation: 1) Generate a sentence containing a numeric expression written down as number words, 2) Generate audio, 3) Convert the sentence such that the numeric expression is written using numeric literals.}
  \label{fig:augmentation}
\end{figure}

\begin{table}[t]
\centering
\begin{tabular}{|c|c|c|}
\hline
Set & Utterances & Hours \\
\hline
Training & 2409 & 2.85 \\
Development & 288 & 0.36 \\
Test & 909 & 1.53 \\
\hline
Training-larger & 3637 & 6.86 \\
\hline
\end{tabular}
\vskip 5pt
\captionsetup{width=0.64\linewidth}
\caption{Numeric expression dataset: Number of utterances and hours for the different parts of the generated numeric expressions dataset.} 
\vskip -5pt
\label{tab:data}
\end{table}

\begin{figure}[t]
  \centering
  \boxed{
  \includegraphics[trim=7.4cm 5.6cm 8.5cm 5.1cm,clip,width=0.9\columnwidth,page=2]{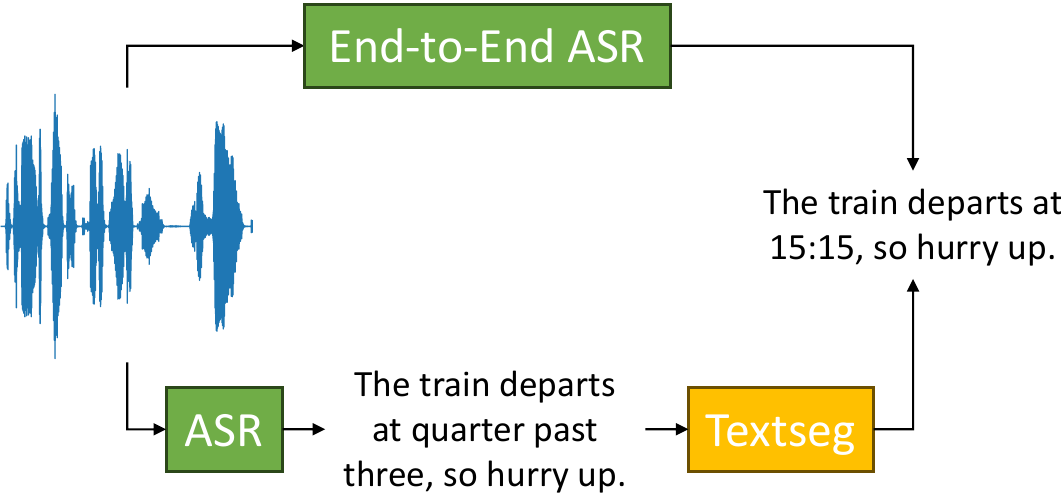}
  }
  \vskip 5pt
  \captionsetup{width=0.92\linewidth}
  \caption{Approaches: Upper: End-to-End ASR model by adapting the ASR model with the numeric expressions dataset. Lower: Cascaded approach by first running the baseline ASR model and then reformatting the output with a text segmentation model.}
  \label{fig:approaches}
\end{figure}

To adapt and test the numeric expression formatting capabilities of our models (see Section \ref{sec:approach}) we created a
\ifcameraready
    \href{https://github.com/chuber11/NumbersTrainingset.git}{numeric expression dataset}
\else
    numeric expression dataset
\fi
(see Figure \ref{fig:augmentation}).

For this, we first used gpt3.5-turbo from OpenAI to generate sentences 
containing the different numeric expression types that we consider written as number words. This is done by a prompt like (the actually used prompt is a little more complex, e.g. to make the LLM not output enumeration):

\begin{quote}
Generate \{n\} diverse [German (optional)] sentences containing a \{numeric expression type\} written down using number words.
\end{quote}

\noindent With half of the executed prompts we generated English sentences and with the other half German sentences.

Then, we used the TTS model tts-1-hd from OpenAI (with voice chosen randomly) to generate audio. For this, it was crucial to have the numeric expressions transcribed as number words since the TTS model did not produce correct output using numeric literals, e.g. \$19.45.

Third, we prompted the LLM to convert the number words into numeric literals in the desired format (see Table \ref{tab:formatting}). This is done by a prompt like

\begin{quote}
Convert the \{numeric expression type\} in the sentences to numeric literals.
\end{quote}

\begin{table*}[t]
\vskip 5pt
\centering
\begin{tabular}{|l||T{0.08\textwidth}|T{0.08\textwidth}|T{0.09\textwidth}|T{0.06\textwidth}|T{0.11\textwidth}|}
\hline
Model & WER (\%) Common voice EN & WER (\%) Common voice DE & \multicolumn{2}{c|}{WER (\%) EN} & WER (\%) DE\\
\cline{4-6}
&&&Numeric expressions & YouTube&Numeric expressions\\
\hline
ASR only & 13.5 & 10.5 & 8.7 & 13.0 & 14.6 \\
ASR+mbart baseline & \textbf{13.4} & \textbf{10.1} & 15.2 & 32.3 & 19.5 \\
ASR+mbart numeric expressions & 13.5 & 10.3 & 5.1 & 17.2 & 9.4 \\
\qquad +more data & \textbf{13.4} & 10.3 & 3.8 & 14.6 & 9.3 \\
ASR+gpt3.5-turbo & 13.5 & 10.6 & 4.3 & \textbf{12.0} & 8.0 \\
ASR+gpt4-turbo & 13.5 & 10.4 & 3.5 & 12.6 & 7.4 \\
ASR+gpt-4o & 13.6 & 10.5 & 3.4 & 12.9 & 7.4 \\
fine-tuned ASR & 13.7 & 10.7 & 3.8 & 12.9 & \textbf{5.7} \\
\qquad +more data & 13.7 & 10.9 & \textbf{3.3} & 13.2 & 5.8 \\
\hline
\end{tabular}
\vskip 5pt
\captionsetup{width=0.82\textwidth}
\caption{Results: WER ($\downarrow$) for English, German on a filtered version of Common voice not containing numeric expressions and the numeric expression / YouTube test set for the different approaches.}
\vskip 5pt
\label{tab:res}
\end{table*}

\begin{table*}[t]
\centering
\begin{tabular}{|l||T{0.045\textwidth}|T{0.06\textwidth}|T{0.045\textwidth}|T{0.06\textwidth}|T{0.045\textwidth}|T{0.06\textwidth}|T{0.045\textwidth}|T{0.06\textwidth}||T{0.045\textwidth}|T{0.06\textwidth}|}
\hline
Model & \multicolumn{2}{l|}{\makecell{Accuracy (\%)\\ years}} & \multicolumn{2}{l|}{\makecell{Accuracy (\%)\\ timestamps}} & \multicolumn{2}{l|}{\makecell{Accuracy (\%)\\ currency amounts}} & \multicolumn{2}{l||}{\makecell{Accuracy (\%)\\ quantities}} & \multicolumn{2}{l|}{\makecell{Accuracy (\%)\\ average}} \\
\cline{2-11}
&Num. exp.&YouTube&Num. exp.&YouTube&Num. exp.&YouTube&Num. exp.&YouTube&Num. exp.&YouTube\\
\hline
ASR only & 97.4 & \textbf{97.9} & 3.4 & 0.0 & 36.3 & 84.3 & 93.3 & 86.0 & 57.6 & 69.5\\
ASR+mbart baseline & 74.3 & 29.7 & 1.0 & 0.0 & 4.4 & 4.6 & 8.6 & 4.6 & 22.1 & 10.2\\
ASR+mbart num. exp. & 96.6 & 90.9 & 20.3 & 2.5 & 71.1 & 76.4 & 91.4 & 64.2 & 69.9 & 61.2\\
\qquad +more data & 95.8 & 91.7 & 39.2 & 9.0 & 73.3 & 78.8 & 93.3 & 74.6 & 75.4 & 65.7\\
ASR+gpt3.5-turbo & 96.6 & 97.7 & 64.9 & 52.3 & 79.3 & 86.1& 96.3 & 86.0 & 84.3 & 81.6\\
ASR+gpt4-turbo & 95.5 & 93.2 & 78.4 & \textbf{92.6} & 94.8 & 80.1 & \textbf{99.4} & 80.1 & 92.0 & \textbf{86.4}\\
ASR+gpt-4o & 95.5 & 94.3 & \textbf{83.8} & 92.3 & 95.6 & 77.4 & \textbf{99.4} & 82.7& \textbf{93.6} & 86.3\\
fine-tuned ASR & \textbf{97.7} & 97.1 & 63.9 & 29.4 & 97.0 & \textbf{89.6} & \textbf{99.4} & 86.0 & 89.5 & 77.4\\
\qquad +more data & \textbf{97.7} & \textbf{97.9} & 75.9 & 39.3 & \textbf{98.5} & 86.6 & \textbf{99.4} & \textbf{89.3} & 92.9 & 79.6\\
\hline
\end{tabular}
\vskip 5pt
\captionsetup{width=0.97\textwidth}
\caption{Results: Accuracy ($\uparrow$) for years, timestamps, currency amounts and quantities on our numeric expression / YouTube test data for the different approaches.}
\vskip -5pt
\label{tab:res2}
\end{table*}

\noindent The output of this step is used as labels for the utterances.

To get a high quality dataset we applied some filtering using simple rules, e.g., output sentences of the third step not containing numeric literals were ignored.

The created data was then divided into training, development and test sets. We noticed that the numeric expressions created by the LLM sometimes are repeated. Therefore, we split the data so that the numeric expressions contained in the three sets were chosen to be pairwise disjoint (see Table \ref{tab:data}). The test set was then read by human annotators to collect real audio samples.

We noticed that for the end-to-end approach (see Section \ref{seq:res}) the performance on the timestamps was worse than the cascaded approaches.
Therefore, we evaluated whether more data generation could help and created more training data (Training-larger) containing timestamps with gpt-4o. This is done by a prompt like

\begin{quote}
Generate a [German (optional)] sentence containing the timestamp \{timestamp\} written down using number words.
\end{quote}

\noindent For \{timestamp\} we iterate over many possibilities, e.g. for English one o'clock, quarter past one, half past one, quarter to one, two minutes past one, two minutes to one. For German, we used equivalent translations. The second and third steps of the data generation are the same as before.


Furthermore, we created a real-world dataset containing numeric expressions by filtering 7,762 hours of human-corrected english YouTube data using regular expressions. Then, we manually corrected errors. The resulting dataset denoted YouTube testset contains 3.0 hours of audio and a total of 1447 numeric expressions.

To evaluate the general performance of our model, we report the WER on the Common voice \cite{ardila2019common} test sets in English and German. We filtered the test sets by excluding utterances containing numeric expressions because the numeric expressions contained in the labels are written down using number words, and we tuned our models to output numeric literals. The English and German test sets contain 2,000 utterances and 3.3 hours of audio each.

\subsection{Models and Approaches}
\label{sec:approach}

We compare cascaded and end-to-end approaches for the formatting of numeric expressions (see Figure \ref{fig:approaches}). For the cascaded approach, we use a trained ASR model and reformat the output using a text segmentation model. For the end-to-end approach, we adapt the trained ASR model by fine-tuning on the training set of our numeric expressions dataset (see Section \ref{sec:data}) with a small learning rate of $10^{-6}$.

We use Whisper \cite{radford2023robust} (whisper-large-v2) as our baseline ASR model and for the text segmentation model we compare using a mbart-based model \cite{liu2020multilingual} (mbart-large-50) and LLMs (gpt3.5-turbo, gpt4-turbo, gpt-4o).

We adapted the pre-trained Mbart model in two steps since we only have limited data for the second step. First, we fine-tuned it to predict the transcript labels of the Common voice training sets (excluding utterances containing numeric expressions similar to the test sets) given the ASR hypothesis generated by our baseline ASR model. This model we denote by \texttt{mbart baseline}. Second, we fine-tuned the model on the numeric expressions dataset to format numeric expressions correctly. This is done by using the sentences where numeric expressions are written down as number words as input and the corresponding sentences where numeric expressions are written down as numeric literals as labels. We denote this model by \texttt{mbart numeric expressions}. For both steps, we froze the embedding of the model and only trained the parameters of the transformer layers, since this yielded better performance than not freezing it.

As LLM we compared GPT3.5 (gpt-3.5-turbo) or GPT 4 \cite{achiam2023gpt} (gpt4-turbo and gpt-4o) with in-context learning using an example for each type of numeric expression (9 examples in total).

\section{Results}
\label{seq:res}

The results can be seen in table \ref{tab:res} (WER) and table \ref{tab:res2} (accuracy of the different types of numeric expressions). 

We see that the ASR+mbart baseline slightly improves the WER on the Commonvoice test sets due to the learned correction of the ASR hypothesis. However, the performance (WER and accuracy) on the numeric expression / YouTube test sets decreased significantly since the model was not trained to predict numeric literals. 

The model ASR+mbart numeric expressions performs better and outperforms the ASR only model on the numeric expression test sets, while there is not much difference on the Common voice test sets. However, the model struggles to format the timestamps (and currency amounts) correctly, e.g., the ASR hypothesis "The library opens at 10 o'clock, but it's best to arrive early." is converted to "The library opens at 17:00, but it's best to arrive early." This is probably due to the limited amount of numeric expressions data. The same can be seen on the YouTube testset where the performance is even worse than the baseline. Using more data helped a bit, but the accuracy on, e.g., timestamps is still less than 40\%/10\%.

Using an LLM as a text segmentation model sometimes (gpt3.5-turbo: 9.7\%, gpt4-turbo: 1.4\%, gpt-4o: 2.7\%) does not follow the prompt, for example, when the input sentence is a question, it is answered. This leads to a completely different transcript and increases in the WER. To circumvent this problem, we compute the WER between input and output of the LLM and if the WER is larger than a threshold (set as 0.5) we ignore to LLM output and return the input instead. 
With this, the LLMs clearly outperform the mbart-based model in terms of both WER and accuracy. The advantage of LLMs is that they were trained on a lot more data. 
Most errors are caused by the LLM not following the prompt, e.g., in the sentence "The bus leaves at five past seven." the timestamps are not changed to 7:05.
Furthermore, the most recently published LLMs generally perform better. Using an LLM which follows the prompt better would probably yield a bit better scores. On the YouTube testset the performance is a bit worse for years (they are sometimes formatted as quantities), currency amounts (e.g. \$9.1 million was replaced by \$9,100,000) and quantities (comma and point are confused) compared to the baseline, however the performance on the timestamps is way better.

It is quite expensive to reformat each hypothesis using an LLM ($\approx$ \$15 to evaluate the ASR+gpt-4o approach on the 4.000 Common voice test sets sentences), especially if the goal is to provide transcription to many customers. Therefore, we experimented with adapting the ASR model end-to-end.
The results show WER performance similar to that of LLM-based approaches. However, the WER on the German numeric expression test set is a bit better. The improvement is due to the fact that for most of the timestamp data the baseline ASR model outputs, e.g. for an audio containing "Ich habe bis fünfzehn Uhr fünfundvierzig Zeit." a transcript "Ich habe bis 15.45 Uhr Zeit". The conversion by the LLM does not remove the "Uhr", which is counted as an error. The fine-tuned ASR model does not output this "Uhr".
The accuracy of the fine-tuned ASR model with more data on the numeric expressions is better than the approaches ASR+gpt3.5-turbo / gpt4-turbo and only slightly worse than ASR+gpt-4o. On the YouTube numeric expressions the fine-tuned ASR model with more data overall is a bit worse than the cascaded approaches with the LLMs. The largest difference is in the timestamp expressions. For example, the timestamps in the audio containing "[...] you can't focus as well at 4pm as you can at 10am." was transcribed as "4pm" and "10am". The end-to-end approach was not able to learn the mapping to "16:00". Increasing the learning rate did not help because the performance on the Common voice test sets dropped significantly.

With our data generation strategy, it is easy to add more formatting rules, e.g. new currency symbols, by performing more augmenting data and adapting the model.

\subsection{Limitations}

The main limitation for the cascaded approach using an LLM (besides that the needed information has to be in the ASR hypothesis) is the ability of the LLM to follow the prompt correctly. This is expected to be handled even better for newer LLMs getting trained.
For the fine-tuned ASR model, the limitation is to obtain diverse data containing suitable numeric expression formatting.


We also tried to adapt the ASR model using batch weighting \cite{huber2020supervised} and/or a factorization/LoRA-based approach \cite{pham2019very,hu2021lora} together with the common voice training dataset. Although the performance on the Common voice test sets improved, which is expected since the training and test datasets are more similar, the performance on the numeric expression formatting was slightly worse. Furthermore, we tried freezing the encoder or only adapting the final projection layer during the adaptation with the numeric expressions data. For both, the performance on the numeric expression test data was slightly worse compared to not freezing any weights.

\section{Conclusion}

In this paper, we tackled the problem of correctly formatting numeric expressions in ASR transcripts.
Our experiments revealed that LLMs, particularly the latest models, deliver strong performance in recognizing and formatting numeric expressions.
However, end-to-end models adapted with synthetic ASR data provide competitive performance.

\section{Acknowledgment}

\ifcameraready
This research was supported in part by a grant from Zoom Video Communications, Inc.
Furthermore, the projects on which this research is based were funded by the Federal Ministry of Education and Research (BMBF) of Germany under the number 01EF1803B (RELATER),
the Horizon research and innovation program of the European Union under grant agreement No 101135798 (Meetween),
and the KIT Campus Transfer GmbH (KCT) staff in accordance with the collaboration with Carnegie – AI.
The authors gratefully acknowledge the support.
\else
BLIND
\fi


\bibliographystyle{IEEEtran}
\bibliography{mybib}

\end{document}